\documentstyle[12pt,fullpage,subeqn,subeqnarray]{article}
\advance\voffset by 1.2truecm
\advance\hoffset by .7truecm
\def\be{\begin{equation}}
\def\ee{\end{equation}}
\def\beq{\begin{eqnarray}}
\def\eeq{\end{eqnarray}}
\def\ds{\displaystyle{}}
\begin{document}
\thispagestyle{empty}
\begin{flushright}
TIFR/TH/02-21 \\
June 2002 \\
\end{flushright}

\vspace{2em}

\begin{center}
\Large{\bf MECHANISMS OF SUPERSYMMETRY BREAKING IN THE MSSM} \\[1cm]
\large{Probir Roy} \\[1cm]
{\sl Tata Institute of Fundamental Research,} \\ 
{\sl Homi Bhabha Road, Mumbai 400 005, INDIA} \\ [1.5cm]
{\it Invited talk given at the} \\ {\it ``Workshop on High Energy Physics
Phenomenology 7''} \\ {\it Harish-Chandra Research Institute, January
2-14, 2002} \\ [1cm]
\end{center}

\noindent{\bf Preliminary Remarks. Gauge mediated supersymmetry
breaking. Gravity mediated supersymmetry breaking. Anomaly mediated
supersymmetry breaking. Gaugino mediated supersymmetry breaking.
Braneworld supersymmetry breaking. Conclusions.}
\vskip 4ex
\pretolerance=10000

\noindent{$\bullet$} {\bf Preliminary Remarks} \\

This will be a somewhat theoretical review of models and mechanisms
for generating soft explicit supersymmetry breaking terms in the
MSSM. There won't be much signal phenomenology except in a few
illustrative cases. Also, I shall be somewhat antihistorical in first
talking about gauge mediation and then coming to gravity mediation
since my subsequent topics, i.e. AMSB, gaugino mediation as well as
braneworld scenarios, connect more 
naturally with the latter.

Our Lagrangian can be decomposed \cite{one} as
\subequations
\beq
{\cal L}_{\rm MSSM} &=& {\cal L}_{\rm MSSM} + {\cal L}_{\rm SOFT} ,
\label{1a} \\ [2mm]
- {\cal L}_{\rm SOFT} &=& {1\over 2} (M_1 \tilde\lambda_0
\tilde\lambda_0 + M_2 \vec{\tilde\lambda} \cdot \vec{\tilde\lambda} +
M_3 \tilde g^a \tilde g^a + h.c.) + V^{\rm SCALAR}_{\rm SOFT} ,
\label{1b} 
\eeq
\newpage
\beq
V^{\rm SCALAR}_{\rm SOFT} &=& \sum_{\tilde f} \tilde f^\ast_i ({\cal
M}^2_{\tilde f})_{ij} \tilde f_j + (m^2_1 + \mu^2) |h_1|^2 + (m^2_2 +
\mu^2) |h_2|^2 \nonumber \\ [2mm]
&& + (B \mu h_1 \cdot h_2 + h.c.) + {\rm trilinear~\
A~terms}.
\label{1c} 
\eeq
\endsubequations
The sfermion summation in (\ref{1c}) covers all left and right chiral
sleptons and squarks. The other scalars, namely the Higgs doublets
$h_{1,2}$, occur explicitly in the RHS. A direct observable
consequence of (1) is the upper bound \cite{one} on the lightest Higgs mass
\[
m_h < 132 \ {\rm GeV} ,
\]
which is a `killing' prediction of the MSSM.

Though ${\cal L}_{\rm SOFT}$ provides a consistent and adequate
phenomenological description of the MSSM, it is ad hoc and ugly. One
would like a more dynamical understanding of its origin. Supersymmetry has
to be broken and spontaneous breakdown would be an elegant
option. Unfortunately, if this is attempted with purely MSSM fields,
disaster strikes in the form of the Dimopoulos-Georgi sumrule
\cite{one}:
\be
STr \ M^2_{\ell_i} + STr \ M^2_{\nu_i} =0= STr \ M^2_{u_i} + STr \ M^2_{d_i} ,
\label{2}
\ee
where $STr \ M^2_f \equiv m^2_{\tilde f_L} + m^2_{\tilde f_R} - 2 m^2_f$
in terms of physical masses and $i$ is a generation index. Evidently,
(\ref{2}) is absurd since, for each generation, some sparticles are
predicted to be lighter than the corresponding particles in
contradiction with observation. 

The way out of this conundrum is to postulate a hidden world of
superfields $\Sigma$ which are singlets under SM gauge
transformations. Let spontaneous supersymmetry breaking (SSB) take
place at a scale $\Lambda_S$ in this hidden sector and be communicated
to the observable world of superfields $Z$ by a set of messenger
superfields $\Phi$ (Fig. 1) -- characterized by some messenger scale
$M_m$. The induced soft supersymmetry breaking parameters in the
observable sector get characterized by the particle-sparticle mass
splitting $\sim M_s = \Lambda^2_S M^{-1}_m$. The messengers could all
be at the Planck scale (i.e. $M_m = M_{PL}$), but such need not be the
case. 
\begin{figure}[h]
\vspace*{2in}
\includegraphics{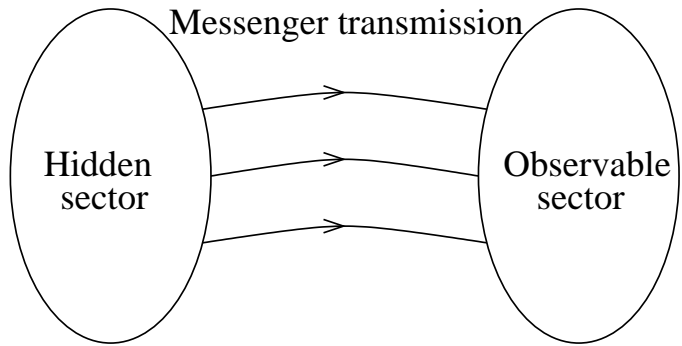}
\caption{The transmission of supersymmetry breaking.}
\end{figure}
They may or may not have nontrivial transformation properties under
the SM gauge group. There are, in fact, two broad categories of
messenger mechanisms: (1) gauge mediation and (2) gravity
mediation. In (1) the messengers are intermediate mass $(\geq$ 100
TeV) fields with SM gauge interactions. In (2) they are near Planck
scale supergravity fields inducing higher dimensional supersymmetry
breaking operators suppressed by powers of $M^{-1}_{PL}$.

\vspace{1em}

\noindent{$\bullet$} {\bf Gauge mediated supersymmetry breaking
\cite{two,three,four}} \\

The messenger superfields here have all the MSSM gauge
interactions. MSSM superfields, with identical gauge interactions but
different flavours, are treated identically by the messengers; thus
there are no FCNC amplitudes. Loop diagrams induce the explicit soft
supersymmetry breaking terms in the MSSM. Loop diagrams, generating
gaugino and scalar masses, are shown in Figs. 2a and 2b with
$\{\phi, \chi\}$ and $\{Z, \psi\}$ being components of $\Phi$ and
$Z$ respectively. Let $S$ be a generic hidden sector chiral superfield
and $\{\Phi_i, \bar\Phi_i\}$ a set messenger chiral
superfields\footnote{$\Phi_i$ and $\bar\Phi_i$ together form a
vectorlike representation of $SU(5)$}, interacting via couplings
$\lambda_i$ in the superpotential 
\be
W_{mess} = \sum_i \lambda_i S \Phi_i \bar\Phi_i .
\label{3}
\ee
SSB in the hidden sector is characterized by the auxiliary component
VEV $\langle F_S \rangle$. A typical messenger mass is given by $M_m
\sim | \lambda_i \langle S \rangle|$. Define
\be
x_i \equiv \left| {\langle F_S \rangle \over \lambda_i \langle S
\rangle} \right| , \ \Lambda \equiv {|\langle F_S \rangle| \over
|\langle S \rangle|} ,
\label{4}
\ee
i.e. $M_m = \Lambda / x_i$. One can then show from the required
positivity of the lowest eigenvalue of the messenger scalar mass
matrix that $0 < x_i < 1$. 
\begin{figure}[h]
\vspace*{1.5in}
\includegraphics{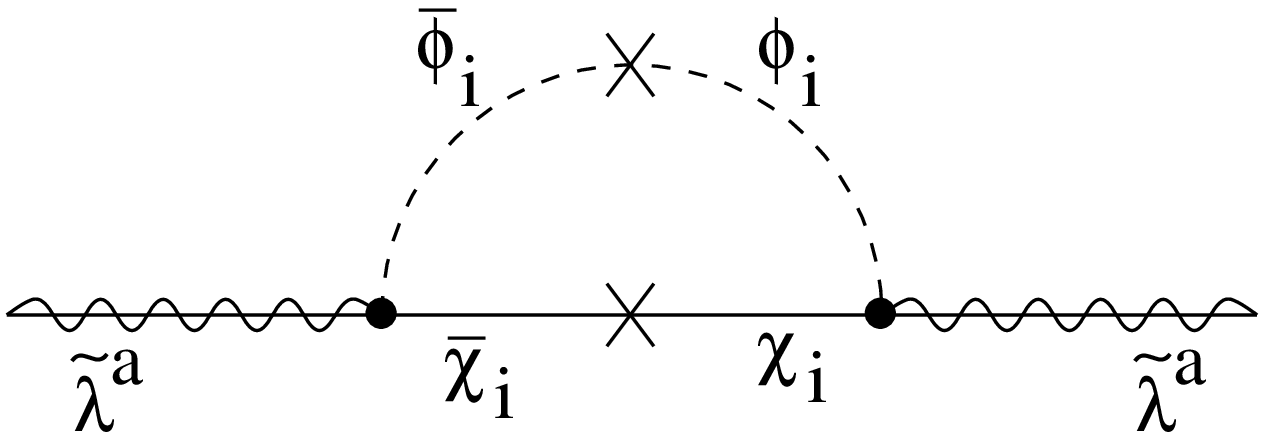}
\includegraphics{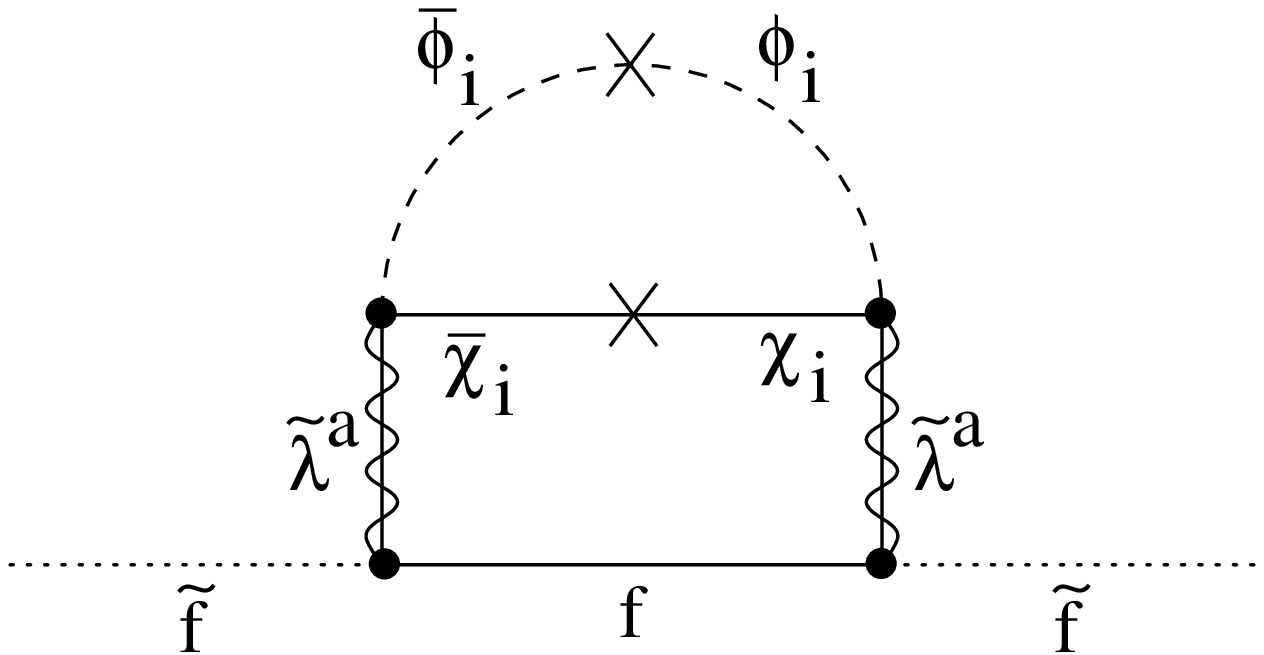}
\caption{The origin of (a) gaugino and (b) scalar masses in GMSB.}
\end{figure}

Gaugino (scalar) masses originate in one (two) loop(s) in the manner
of Fig. 2a(b):
\subequations 
\beq
M_\alpha &=& (g^2_2 / 16 \pi^2) \Lambda \sum_\alpha 2 T_\alpha (R_i) g
(x_i) , \label{5a} \\ [2mm]
m^2_{\tilde f, h} &=& 2 \Lambda^2 \sum_\alpha (g^2_2 / 16 \pi^2)^2
C_\alpha \sum_i 2 T_\alpha (R_i) f (x_i) .
\label{5b}
\eeq
\endsubequations
Here Tr $T^a (\phi_i) T^b (\phi_i) = T_\alpha (R_i) \delta^{ab}$
where the trace is over the representation $R_i$ of $\phi_i$ in the
gauge group factor $G_\alpha$ and $C_\alpha$ is the quadratic Casimir 
$(\sum_a T^a T^a)_{G_\alpha}$ of the latter. Moreover, 
\subequations
\beq
g(x) &=& x^{-2} \left[ (1+x) \ln (1+x) - (1-x) \ln (1-x)
\right], \label{6a} \\ [2mm]
f(x) &=& x^{-2} (1+x) \left[ \ln (1+x) - 2 Li_2 \left({x \over
1+x}\right) + {1\over 2} Li_2 \left({2x \over 1+x}\right)\right] + (x
\leftrightarrow -x) , 
\label{6b}
\eeq
\endsubequations
$Li_2$ being the dilogarithm. The behaviour of $g(x)$ and $f(x)$ in
the region $0 \leq x \leq 1$ is shown in Fig. 3. They are practically
unity for a large range of $x$. In this situation $\sum_\alpha 2
T_\alpha (R_i)$ factorizes and becomes $n_5$ for $SU(3)_C$ or
$SU(2)_L$ but $\sum_i (Y_i/2)^2 = {5\over 3} n_5$ for $U(1)_Y$, 
where $n_5$ is the number of complete {\bf 5} $\oplus$
{\bf 5} messenger representations of $SU(5)$. Now one can write 
\subequations
\beq
M_\alpha &\simeq& (g^2_\alpha / 16 \pi^2) n_5 \Lambda \label{7a} \\ [2mm]
m^2_{\tilde f, h} (M_m) &\simeq& 2 n^{-1}_5 \left[ C_3 M^2_3
(M_m) + C_2 M^2_2 (M_m) + {3\over 5} \left({Y\over 2}\right)^2 M^2_1 (M_m)\right],
\label{7b}
\eeq
\endsubequations
\begin{figure}
\vspace*{2.1in}
\includegraphics{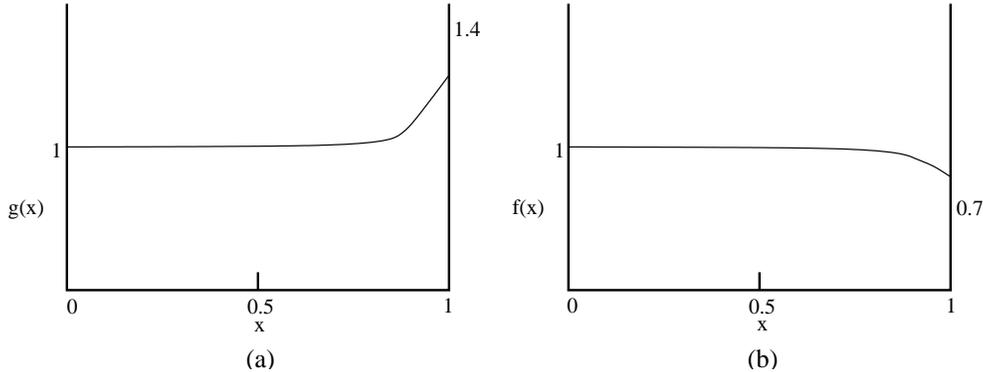}
\caption{The behaviour of (a) $g(x)$ and (b) $f(x)$}
\end{figure}
where $C_3 = {4\over 3} \ (0)$ for an $SU(3)_C$ triplet (singlet) and
$C_2 = {3\over 4} \ (0)$ for an $SU(2)_L$ doublet (singlet).  To one
loop the
gaugino masses (\ref{7a}) vary with RG evolution in the same way as
$g^2_\alpha$, while the scalar masses (\ref{7b}) are specified at an
energy scale $M_m$ corresponding to messenger masses. The trilinear
coupling $A$
parameters get induced at the two loop level and can be taken to
vanish at the scale $M_m$ --- becoming nonzero at lower energies via RG
evolution. The parameters $\mu, B$ are kept free to implement the
radiative EW breakdown mechanism, the validity of which implies the
bounds \cite{four}
\be
50 \ {\rm TeV} < M_m < \sqrt{n_5} \times 10^{14} \ {\rm GeV} .
\label{8}
\ee

The minimal GMSB model, called mGMSB, is characterized by the
parameter set
\be
\{p\} = \{\Lambda, M_m, \tan\beta, n_5, sgn \mu \} .
\label{9}
\ee
Linear RG interpolation of sfermion squarel masses from the boundary
values of (\ref{7b}) at the scale $M_m$ to lower energies $\sim
\Lambda$ yield, with $t_M = \ln M_m/\Lambda$, the one loop expressions
\newpage
\subequations
\beq
m^2_{\tilde e_R} (100 \ {\rm GeV}) &=& M^2_1 (100 \ {\rm GeV})
\left[1.54 n^{-1}_5 + 0.05 + (0.072 n^{-1}_5 + 0.01)
t_M\right] \nonumber \\ [2mm] 
&& \hspace{8cm} + s^2_W D , \label{10a} \\ [2mm]
m^2_{\tilde e_L} (100 \ {\rm GeV}) &=& M^2_2 (100 \ {\rm GeV}) \left[1.71
n^{-1}_5 + 
0.11 + (0.023 n^{-1}_5 + 0.02) t_M\right] \nonumber \\ [2mm]
&& \hspace{6.7cm} + (0.5 - s^2_W) D , \label{10b} \\ [2mm] 
m^2_{\tilde v} (100 \ {\rm GeV}) &=& M^2_2 (100 \ {\rm GeV})
\left[1.71 n^{-1}_5 + 
0.11 + (0.023 n^{-1}_5 + 0.02) t_M\right] \nonumber \\ [2mm]
&& \hspace{8cm} - 0.5 D , \label{10c} \\ [2mm]
m^2_{\tilde u_L} (500 \ {\rm GeV}) &=& M^2_3 (500 \ {\rm GeV})
\left[1.96 n^{-1}_5 + 
0.31 + (- 0.102 n^{-1}_5 + 0.037) t_M\right] \nonumber \\ [2mm]
&& \hspace{6.5cm} - (0.5 - 0.66 s^2_W) D , \label{10d} \\ [2mm] 
m^2_{\tilde d_L} (500 \ {\rm GeV}) &=& M^2_3 (500 \ {\rm GeV})
\left[1.96 n^{-1}_5 + 
0.31 + (- 0.102 n^{-1}_5 + 0.037) t_M\right] \nonumber \\ [2mm]
&& \hspace{6.5cm} + (0.5 - 0.66 s^2_W) D , \label{10e} \\ [2mm]
m^2_{\tilde u_R} (500 \ {\rm GeV}) &=& M^2_3 (500 \ {\rm GeV})
\left[1.78 n^{-1}_5 + 
0.30 + (- 0.103 n^{-1}_5 + 0.035) t_M\right] \nonumber \\ [2mm]
&& \hspace{8cm} - 0.66 s^2_W D , \label{10f} \\ [2mm] 
m^2_{\tilde d_R} (500 \ {\rm GeV}) &=& M^2_3 (500 \ {\rm GeV})
\left[1.77 n^{-1}_5 + 
0.30 + (- 0.103 n^{-1}_5 + 0.034) t_M\right] \nonumber \\ [2mm]
&& \hspace{8cm} + 0.33 s^2_W D , 
\label{10a} 
\eeq
\endsubequations
where $s^2_W \equiv \sin^2 \theta_W$ and $D \equiv - M^2_Z \cos 2
\beta$. This sfermion mass spectrum may look like that in $m$SUGRA in
the limit when $m_0 \ll M_{1/2}$. But that limit in $m$SUGRA is ruled
out by the required absence of charge and colour violating vacua, as
will be pointed out later. Thus
the contents of the sfermion mass spectrum, specifically the squark to
slepton and singlet to doublet sfermion mass rations, distinguish
mGMSB. A final point on scalar masses is that the magnitude of the
$|\mu|$ parameter is 
forced to become large by the requirement of EW symmetry breakdown: 
\be
|\mu| \geq {2\over 3} n^{-1}_5 M_3 (M_m) .
\label{11}
\ee
Such a large $|\mu|$ makes the CP even charged (heavy neutral) Higgs
$H^\pm (H)$ as well as the CP odd neutral Higgs $A$ very heavy and
tightens the upper bound of 132 GeV on $h$ in general MSSM to
\be
m_h < 120 \ {\rm GeV} .
\label{12}
\ee

The gravitino mass is given by
\[
m_{3/2} = \sqrt{{1\over 3}} {|\langle F_{_S}\rangle| \over M_{PL}} = O
({\rm keV}) .
\]
Thus the gravitino behaves here like an ultralight pseudo-Goldstino
and is the LSP. If $\tilde\chi^0_1$ is the NLSP, it will have decays like
$\tilde\chi^0_1 \to \gamma \tilde G, Z\tilde G, h\tilde G$ etc. One
can estimate that
\be
\tau_{\rm NLSP} \geq 6 \times 10^{-14} \left({100 \ {\rm GeV} \over
M_{\tilde\chi^0_1}}\right)^5 \left[{\Lambda M_m \over (64 \lambda \
{\rm  TeV})^2}\right]^2 {\rm secs} 
\label{13}
\ee
and $c_{\tau \rm NLSP}$ will be less than the length dimension of a
detector if $M_m > 50 \ {\rm TeV}$. The decay photon for the $\gamma
\tilde G$ final state provides a characteristic
signature. Another interesting possibility is that of $\tilde\tau_1$
being the NLSP in which case one will have the prompt decay
$\tilde\tau_1 \to \tilde G\tau$ and a hard, isolated $\tau$ in
addition to large $E\!\!\!\!/_T$ and leptons and/or jets from
cascades will be a distinctive GMSB signal.

The GMSB scenario suffers from a severe finetuning problem between 
$|\mu|$ and $|\mu B|$. Eq.(\ref{11}) makes $|\mu|$ quite large. The
$\mu$ parameter originates in the GMSB scenario from a term
$\lambda_\mu S H_1\!\cdot\!H_2$ in the superpotential and a VEV
$\langle s \rangle$ for the scalar component of $S$, but that leads to
the soft $B\mu$ term in eq. (\ref{1c}) also. Then consistency with
eq.(\ref{11}) requires $|B| > 30$ TeV, which is rather large and bad
for the finetuning aspect in the stabilization of the weak scale.\\

\noindent{\bf $\bullet$ Gravity mediated supersymmetry breaking} \\

The messengers in this scenario \cite{five} are the superfields of an
$N=1$ supergravity theory, coupled to matter, with the messenger mass
scale being close to the Planck scale. It has two major advantages:
(1) the presence of gravity in local supersymmetry is utilized
establishing a connection between global and local supersymmetry; (2)
the theory automatically contains operators which can transmit
supersymmetry breaking from the hidden to the observable sector. There
are two disadvantages, though. First, since $N=1$ supergravity theory
is not renormalizable, one has to deal with an effective theory at
sub-Plankian energies vis-a-vis poorly understood Planck scale
physics. In particular, naive assumptions, made to
simplify the cumbrous structure of this theory, may not hold in
reality. Second, there are generically large FCNC effects of the form
\be
{\cal L}_{eff} \sim \int d^4 \phi \ h M^{-2}_{PL} (\Sigma^+ \Sigma \
Z^+ Z) , 
\label{14}
\ee
$h$ being a typical Yukawa coupling strength. \\

\noindent{\it Lightning summary of $N=1$ supergravity theory} \\

The general supergravity invariant action, with matter superfields
$\Phi_i$, gauge superfields $V= V^a T^a$ and corresponding spinorial
field-strength superfields $W^a$, is \cite{one,five}.
\be
S = \int d^6 Z \left[ - {1\over 8} {\cal D} {\cal D} {\cal K} \{
(\Phi^\dagger e^V)_i. \Phi_j\} + {\cal W} (\Phi_i) + {1\over 4} f_{ab}
(\Phi_i) W^{aA} W^b_A \right] + h.c. 
\label{15}
\ee
Here ${\cal W}$ is the superpotential, $f_{ab} (\Phi_i)$ an unknown
analytic function of $\Phi$ and ${\cal K}$ an unknown hermition
function. The definition
\be
{\cal G} \equiv M^2_{PL} \left[ - 3 \ln \ \{ - {1\over 3} M^{-2}_{PL}
{\cal K} (\Phi^\dagger e^V, \Phi)\} - \ln \ \{ M^{-6}_{PL} | {\cal W}
(\Phi)|^2 \}\right] 
\label{16}
\ee
and Weyl rescaling \cite{one,four} enable us to rewrite the non-KE
terms in the integrand of eq.(\ref{15}) as the potential
\be
V = - F_i {\cal G}^i_j \bar F^j - 3 M^4_{PL} e^{-{\cal G}/M^2_{PL}} +
{1\over 2} \sum_\alpha g^2_\alpha D^{\alpha a} D^{\alpha a} ,
\label{17}
\ee
with 
\subequations
\be
F_i = M_{PL} e^{- {\cal G}/(2M^2_{PL})} ({\cal G}^{-1})^j_i {\cal G}_j
+ {1\over 4} f^\ast_{ab,k} ({\cal G}^{-1})^k_i \bar\lambda^a
\bar\lambda^b - ({\cal G}^{-1})^k_i {\cal G}^{jL}_k \chi_j \chi_i ,  
\label{18a}
\ee
\be
D^{\alpha a} = {\cal G}^i (T^{\alpha a})^j_i \phi_j ,
\label{18b}
\ee
\endsubequations
$G_\alpha$ being the $\alpha$th factor of the gauge group
$G = \prod_\alpha G_\alpha$. 

The separation between the hidden sector superfields $\Sigma$ and the
observable sector ones $Z_i$ is effected by writing
\[
\Phi_i \equiv \{ Z_i, \Sigma \}, \ \phi_i \equiv \{ z_i, \sigma \} , \
\bar\Phi^i \equiv \{ \bar z^i, \bar\sigma\}
\]
and assuming the additive split of the superpotential into observable
and hidden parts
\be
{\cal W} (\Phi_i) = {\cal W}_0 (Z_i) + {\cal W}_h (\Sigma) .
\label{19}
\ee
The spontaneous breakdown of supersymmetry in the hidden sector can be
implemented through either a nonzero VEV $\langle F_\Sigma \rangle$ of
an auxiliary 
component of the $\Sigma$ superfield or a condensate $\langle
\lambda_\Sigma \lambda_\Sigma \rangle$ of hidden sector gauginos. As a
result, the gravitino becomes massive through the super-Higgs
mechanism: $m_{3/2} = M_{\rm PL} e^{-<{\cal G}>/(2M^2_{\rm
PL})}$. Furthermore, soft supersymmetry breaking parameters $A_{ijk}$
and $B$ are generated in the observable sector with magnitudes $\sim
\langle F_\Sigma \rangle/M_{\rm PL}$ or $\langle \lambda_\Sigma
\lambda_\Sigma \rangle/M^2_{\rm PL}$. Scalar and gaugino masses are
also generated respectively as \cite{one,four} 
\subequations
\beq
m_i &=& O (m_{3/2}) , \label{20a} \\ [2mm]
M_{ab} &=& {1\over 2} m_{3/2} \langle {\cal G}^l ({\cal G}^{-1})^k_l
f^\ast_{ab, k} \rangle . 
\label{20b}
\eeq
\endsubequations
The procedure suggested in Ref.\cite{six} was to use these results as
boundary conditions at the unification scale $M_U$, where $M_W \ll M_U
< M_{\rm PL}$, and evolve down to laboratory energies by RG
equations. \\

\noindent{\it mSURGA and beyond} \\

mSUGRA is a model characterized by the following specific boundary
conditions on soft supersymmetry breaking parameters at the unifying
scale $M_U$:
\begin{itemize}
\item{} universal gaugino masses $M_\alpha (M_U) = M_{1/2},
\ \ \forall \alpha$,
\item{} universal scalar masses $m^2_{ij} (M_U) = m^2_0
\ \delta_{ij}$,
\item{} universal trilinear scalar couplings $A_{ijk} (M_u) = A_0 \ \ 
\forall i, j, k$ .
\end{itemize}
The soft supersymmetry breaking parameters are then treated as
dynamical variables evolving from their boundary values via RG
eqns. The complete set of parameters needed for mSUGRA is
\be
\{p\} = (sgn \ \mu, \ m_0 , \ M_{1/2} , \ A_0 , \ \tan \beta ) .
\label{21}
\ee
The magnitude $|\mu|$ of the higgsino mass gets fixed by the
requirement of the EW symmetry breakdown. Among some of the immediate
consequences are the predicted gaugino mass ratios at electroweak
energies 
\be
M_3 (100 \ {\rm GeV}) : M_2 (100 \ {\rm GeV}) : M_1 (100 \ {\rm GeV})
\simeq 7 : 2 : 1 
\label{22}
\ee
and the interpolating sfermion mass formulae
\subequations
\beq
m^2_{\tilde l_R} (100 \ {\rm GeV})  &=& m^2_0 + 0.15 M^2_{1/2} - s^2_W
M^2_Z \cos 2 \beta , \label{23a} \\ [2mm]
m^2_{\tilde l_L} (100 \ {\rm GeV})  &=& m^2_0 + 0.53 M^2_{1/2} +
(T^l_{3L} - Q_l s^2_W) M^2_Z \cos 2 \beta , \label{23b} \\ [2mm]
m^2_{\tilde q_L} (500 \ {\rm GeV})  &=& m^2_0 + 5.6 M^2_{1/2} +
(T^q_{3L} - Q_q s^2_W) M^2_Z \cos 2 \beta , \label{23c} \\ [2mm]
m^2_{\tilde u_R} (500 \ {\rm GeV})  &=& m^2_0 + 5.2 M^2_{1/2} +
{2\over 3} s^2_W M^2_Z \cos 2 \beta , \label{23d} \\ [2mm]
m^2_{\tilde d_R} (500 \ {\rm GeV})  &=& m^2_0 + 5.1 M^2_{1/2} -
{1\over 3} s^2_W M^2_Z \cos 2 \beta . 
\label{23e}
\eeq
\endsubequations
Let us make two final remarks on mSUGRA. First, the required absence of charge
and colour violating minima disallows \cite{seven} the limit $m_0 \ll
M_{1/2}$ for mSUGRA, thereby establishing its mutual exclusivity
vis-a-vis the mGMSB spectrum. Second, the $\mu$-term is somewhat less
of a problem here than in GMSB since something like the
Giudice-Masiero mechanism \cite{eight} for generating it can be
incorporated within this framework.

Going beyond mSUGRA, one sometimes pursues a constrained version of
the MSSM, called CMSSM, where the radiative EW symmetry breakdown
condition is not insisted upon. Moreover, separate universal masses
are assumed at $M_U$ for fermions and Higgs bosons, since they
supposedly belong to different representations of the GUT group. Now
the parameter set is expanded to 
\be
\{ p \}_{CMSSM} = \{ \mu, \ m_A, \ m_{\tilde f}, \ M_{1/2}, \ A_0, \
\tan\beta \} .
\label{24}
\ee
Further, the spectrum plus associated phenomenology get related to but
remain somewhat different from those in mSUGRA in having less
predictivity. A basic criticism is the lack of justification for the
still present subset of universality assumptions at $M_U$. But one is
beset with severe FCNC problems if these are discarded. In particular,
near mass degeneracy is needed for squarks of the first two
generations and the same goes for sleptons.

There have been attempts to avoid such ad hoc universality assumptions
and instead forbid FCNC through some kind of a family symmetry. A
spontaneously proken $U(2)_F$, with doublets $L_a, R_a \ (a = 1, 2)$ and
singlets $L_3, R_3$, has been invoked for this purpose
\cite{nine}. The scheme works provided additional Higgs fields are
introduced. Specifically, one needs `flavon' fields $\phi^{ab}$ that
are antisymmetric in $a, b$ and have the VEV $\langle \phi^{ab} \rangle = v 
\epsilon^{ab} = \pmatrix{0 & v \cr - v & 0}$. \\

\noindent{\bf Anomaly mediated supersymmetry breaking} \\

This is a scenario \cite{ten} in which the FCNC problem is naturally
solved and yet many of the good features of usual gravity mediation
are retained. It makes use of three branes, which are three
dimensional stable solitonic solutions (of the field equations)
existing in a bulk of higher dimensional spacetime --- originally
discovered in String Theory. Consider two parallel three branes, one
corresponding to the observable and the other to the hidden
sector. This means that all matter and gauge superfields belonging to
one sector are pinned to the corresponding brane. The two branes are
separated by a bulk distance $r_c \sim$ compactification radius. Only
gravity propagates in tbe bulk. Any direct exchange between the two
branes, mediated by a bulk field of mass $m$, say, will be suppressed
in the amplitude by the factor $e^{-mr_c}$. (One assumes that there
are no bulk fields lighter than $r_c^{-1}$). SUGRA fields, propagating
in the bulk, get eliminated by the rescaling transformation $SZ \to Z$
where $S$ is a compensator left chiral superfield. However, this
rescaling transformation is anamolous, giving rise to a loop induced
superconformal anomaly which communicates the breaking of supersymmetry
from the hidden to the observable sector. Being topological in origin,
it is independent of the bulk distance $r_c$ and is also flavour blind. In
consequence, there is no untowardly induction of FCNC amplitudes. One
obtains one loop gaugino masses and two loop squared scalar masses as
under 
\begin{figure}[h]
\vspace*{2in}
\includegraphics{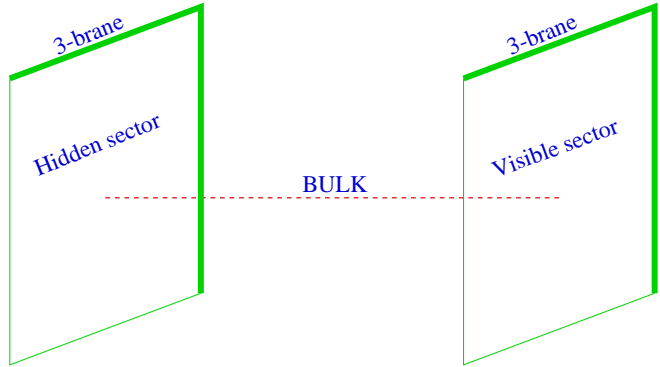}
\caption{Hidden and observable branes in the bulk}
\end{figure}
\subequations
\beq
M_\alpha &=& M {\beta (g_\alpha) \over g_\alpha} , \label{25a} \\ [2mm]
m^2_i (Q) &=& - {1\over 4} \left[ \beta (g_\alpha) {d\gamma_i \over
dg_\alpha} + \beta_\gamma {\partial\gamma_i \over \partial Y} \right]
m^2_{3/2}.
\label{25b}
\eeq
\endsubequations
Here $Y$ is a generic Yukawa coupling strength while $\gamma_i$ is the
anomalous dimension of the ith matter superfield (N.B. $\gamma_{ij} = \gamma_i
\delta_{ij}$). In addition, the trilinear A-couplings are given by
\be
A_{ijk} = - {1\over 2} (\gamma_i + \gamma_j + \gamma_k) .
\label{26}
\ee
An interesting fallout of eq.(25a) is the numerical proportionality 
\be
M_1 (100 \ {\rm GeV}) : M_2 (100 \ {\rm GeV}) : M_3 (100 \ {\rm GeV})
= 2.8 : 1 : 7.1 .
\label{27}
\ee
However, eq. (25b) leads to the disastrous consequence of physical
sleptons becoming tachyonic since it implies $m^2_{sleptons} (M_W) <
0$. 

Various strategies have been attempted to evade the tachyonic slepton
problem mentioned above. The simplest procedure, which defines the
mAMSB model, is to add a universal dimensional constant $m^2_0$ to
$m^2_i$. The manifest RG invariance of eq. (25b) is lost now and one
obtains
\subequations
\beq
m^2_i &=& C_i (16\pi^2)^{-2} m^2_{3/2} + m^2_0 , \label{28a} \\ [2mm]
A_{t, b, \tau} &=& (16 \pi^2)^{-1} m_{3/2} h^{-1}_{t, b, \tau}
\hat\beta_{h_{t,b,\tau}} , 
\label{28b}
\eeq
\endsubequations 
where the $\hat\beta$'s and the $C_i$'s are given in Table 1. 
The main
spectral feature in the bosino sector of this model is that the
lightest meutralino $\tilde\chi^0_1$ and the lightest chargino
$\tilde\chi^\pm_1$ are nearly mass degenerate, both being winolike,
while the next higher neutralino $\tilde\chi^0_2$ is somewhat
heavier. As a result, $\tilde\chi^\pm_1$ is longlived and can be
observed \cite{eleven} if
\[
180 \ {\rm MeV} \ < M_{\tilde\chi^\pm_1} - M_{\tilde\chi^0_1} < 1 \
{\rm GeV} .
\]
The left selectron $\tilde e_L$ is also nearly mass degenerate with
the right selectron $\tilde e_R$. 
\begin{table}[h]
\begin{center}
\begin{tabular}{|lcl|} \hline 
$\hat\beta_{h_t}$ &=& $h_t \left( - {\ds{13\over 15}} g^2_1 - 3 g^2_2 - {\ds{16
\over 3}} g^2_3 + 6 h^2_t + h^2_b \right)$ , \\ [3mm]
$\hat\beta_{h_b}$ &=& $h_t \left( - {\ds{7\over 15}} g^2_1 - 3 g^2_2 -
{\ds{16 \over 3}} g^2_3 + h^2_t + 6h^2_b + h^2_\tau \right)$ , \\ [3mm] 
$\hat\beta_{h_\tau}$ &=& $h_\tau \left( - {\ds{9\over 5}} g^2_1 - 3 g^2_2 +
3h^2_b + 4h^2\tau \right)$ , \\  [3mm]
$C_Q$ &=& $- {\ds{11\over 50}} g^4_1 - {\ds{3\over 2}} g^4_2 + 8 g^4_3 + h_t
\hat\beta_{h_t} + h_b \hat\beta_{h_b}$ ,  \\ [3mm] 
$C_{\bar U}$ &=& $- {\ds{88\over 25}} g^4_1 + 8 g^4_3 + 2h_t
\hat\beta_{h_t}$ ,  \\  [3mm]
$C_{\bar D}$ &=& $- {\ds{22\over 25}} g^4_1 + 8 g^4_3 + 2h_b
\hat\beta_{h_b}$ , \\ [3mm]
$C_L$ &=& $- {\ds{99\over 50}} g^4_1 - {\ds{3\over 2}} g^4_2 + h_\tau
\hat\beta_{h_\tau}$ , \\ [3mm]
$C_{\bar E}$ &=& $- {\ds{198\over 25}} g^4_1 + 2h_\tau
\hat\beta_{h_\tau}$ , \\ [3mm]
$C_{H_2}$ &=& $- {\ds{99\over 50}} g^4_1 - {\ds{3\over 2}} g^4_2 + 3 h_t
\hat\beta_{h_t}$ , \\ [3mm]
$C_{H_1}$ &=& $- {\ds{99\over 50}} g^4_1 - {\ds{3\over 2}} g^4_2 + 3 h_b
\hat\beta_{h_b} + h_\tau \hat\beta_{h_\tau}$ \\ [3mm] \hline  
\end{tabular}
\caption{Expressions for $C_i$'s and $\hat\beta$'s.}
\end{center}
\end{table}
\medskip

\newpage

\noindent{\it Gaugino mediated supersymmetry breaking} \\

In this scenario \cite{twelve}, sometimes called -inoMSB, there are
once again two separated parallel three branes in a higher dimensional
bulk. But now only observable matter superfields are pinned to the
corresponding brane, while gauge and Higgs superfields can propagate
in the bulk. In this situation an interbrane gaugino or higgsino loop
(cf. Fig. 4), in addition to the 
\begin{figure}[h]
\vspace*{2.3in}
\includegraphics{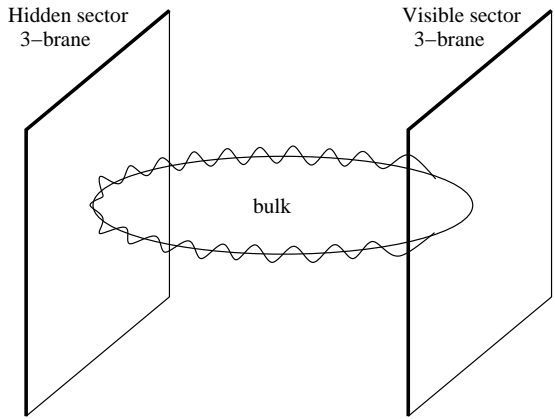}
\caption{An interbrane -inoloop.}
\end{figure}
superconformal anomaly, can transmit supersymmetry breaking from the
hidden to the observable sector. For several three branes, located in
the bulk, the general decomposition of the Lagrangian is 
\be
{\cal L}_D = {\cal L}_{\rm BULK} (\Phi (x, y)) + \sum_j \delta^{(d-4)}
(y - y_j) {\cal L}_j (\Phi (x,y), \chi_j (y)) .
\label{29}
\ee
\noindent In eq.(29) $\Phi (x, y)$ is a typical superfield propagating in the
bulk, whereas $\chi_j (y)$ is a typical superfield localized on the
$j$th brane. This type of a scenario does not seem to have any obvious
problem. On the other hand, it has the following interesting features. 
\begin{itemize}
\item{} $M_{1/2} \sim m_{3/2} \sim |m_{H1}| \sim |m_{H2}| \sim |\mu
B|$. 
\item{} Sleptons are never tachyonic. 
\item{} The $\mu$ problem can be tackled. 
\item{} The near mass degeneracies $M_{\tilde\chi^0_1} \sim
M_{\tilde\chi^\pm_1}, \ \ m_{\tilde e_L} \sim m_{\tilde e_R}$ of mAMSB are
lost. 
\end{itemize}
A sample of sparticle masses for the given input parameters is shown
in Table 2.

\bigskip 

\noindent{\it Braneworld supersymmetry breaking} \\

With two separated and parallel three branes in a higher dimensional
bulk, one can have more general mechanisms for the transmission of
supersymmetry breaking. I just have time to mention them without going
into much detail. One can have scenarios \cite{thirteen}  using
the\break  

$$
\begin{tabular}{|c|c|c|c|c|} \hline
& & Point 1 & Point 2 & Point 3 \\ \hline
inputs: & $M_{1/2}$ & 200 & 400 & 400 \\ 
        & $m^2_{H_u}$ & (200)$^2$ & (400)$^2$ & (400)$^2$ \\
        & $m^2_{H_d}$ & (300)$^2$ & (600)$^2$ & (400)$^2$ \\
        & $\mu$ & 370 & 755 & 725 \\
        & B & 315 & 635 & 510 \\
        & $y_t$ & 0.8 & 0.8 & 0.8 \\ \hline
neutralinos: & $M_{\chi^0_1}$ & 78 & 165 & 165 \\
             & $M_{\chi^0_2}$ & 140 & 315 & 315 \\
             & $M_{\chi^0_3}$ & 320 & 650 & 630 \\
             & $M_{\chi^0_4}$ & 360 & 670 & 650 \\ \hline
charginos: & $M_{\chi^\pm_1}$ & 140 & 315 & 315 \\
           & $M_{\chi^\pm_2}$ & 350 & 670 & 645 \\ \hline
Higgs:     & $\tan\beta$ & 2.5 & 2.5 & 2.5 \\ 
           & $m_{h^0}$ & 90 & 100 & 100 \\ 
           & $m_{H^0}$ & 490 & 995 & 860 \\ 
           & $m_A$ & 490 & 1000 & 860 \\
           & $m_{H^\pm}$ & 495 & 1000 & 860 \\ \hline
sleptons:  & $m_{\tilde e_R}$ & 105 & 200 & 160 \\
           & $m_{\tilde e_L}$ & 140 & 275 & 285 \\
           & $m_{\tilde v_L}$ & 125 & 265 & 280 \\ \hline
stops:     & $m_{\tilde t_1}$ & 350 & 685 & 690 \\
           & $m_{\tilde t_2}$ & 470 & 875 & 875 \\ \hline
other squarks: & $m_{\tilde u_L}$ & 470 & 945 & 945 \\ 
               & $m_{\tilde u_R}$ & 450 & 905 & 910 \\
               & $m_{\tilde d_L}$ & 475 & 950 & 945 \\
               & $m_{\tilde d_R}$ & 455 & 910 & 905 \\ \hline
gluino: & $M_3$ & 520 & 1000 & 1050 \\ \hline
other parameters: & $M_{1/2}$ & 16 & 50 & 50 \\
             & $\mu$ & 19 & 78 & 78 \\ \hline
\end{tabular}
$$
\centerline{Table 2: Sample points in parameter space. All masses are in
GeV. In the first two points,} \centerline{the LSP is mostly Bino, while in the
third it is a right-handed slepton.~}

\bigskip

\noindent 
Randall-Sundrum `warped' metric $ds^2 = e^{-2k|r|} dx^\mu dx^\nu
\eta_{\mu\nu} + dr^2$, with $k$ real and positive, leading to a VEV
$\langle {\cal W} \rangle$ of the superpotential. Alternatively, one could
have compactifications \cite{fourteen} analogous to string
compactifications on the orbifold $S^1/Z_2 \times Z^\prime_2$. A third
possibility \cite{fifteen} is to study general string or Horava-Witten
compactifications of M-theory, yielding two separated three branes in
a bulk. The last approach seems to provide some rationale for R-parity
conservation. Generically, though, these scenarios do {\it not} yield
the kind of K\"ahler potentials required for AMSB or -inoMSB
models. The other phenomenologically interesting approach
\cite{sixteen}, based on string compactifications, is where SUSY
breaking gets mediated by dilatino fields or superpartners of moduli
fields and develops gravity mediated type of a pattern at lower energies. 

\bigskip

\noindent{\bf Conclusion} \\

We can summarize our conclusions in four points. (1) Gauge mediated
supersymmetry breaking has a distinct $\gamma (l) + E\!\!\!\!/_T$ signal,
but suffers from a severe $\mu$ vs $\mu B$ problem. (2) Gravity
mediated supersymmetry breaking can generate the archetypal MSSM at
electroweak energies, but has generic FCNC problems requiring
additional input assumptions; with an extra singlet the $\mu$ problem
can be solved by the Giudice-Masiero mechanism. (3) AMSB has the
advantages of the gravity mediated scenario, but no FCNC problem;
solutions to the tachyonic slepton disaster tend to be ad hoc. (4)
Gaugino/higgsino mediation can lead to a phenomenologically viable
model, free of many of the previous problems, but the required
braneworld scenario does not seem easily derivable from String
Theory. 

\medskip

I compliment the organizers of WHEPP7 for a great workshop and thank
B.C. Allanach and R.M. Godbole for their helpful comments. I also
acknowledge the hospitality of the Department of Physics, University
of Hawaii, where this talk was written up and thank X. Tata for
several clarifying discussions.

\end{document}